\documentclass[prl,twocolumn,showpacs]{revtex4}
\usepackage{amsmath,amssymb,graphicx,color,bm}

\begin{document}

\title{Bosonic Cascade Laser}

\author{T. C. H. Liew$^1$, M. M. Glazov$^{2,3}$,
  K. V. Kavokin$^{2,3}$, I. A. Shelykh$^{1,4}$,
  M. A. Kaliteevski$^{2,5}$, A. V. Kavokin$^{3,6}$}
\address{$^1$ Division of Physics and Applied Physics, Nanyang
  Technological University 637371, Singapore \\
$^2$ Ioffe Phyical-Technical Institute of the RAS, 26,
Polytechnicheskaya, St-Petersburg, Russia \\
$^3$ Spin Optics Laboratory, St-Petersburg State University, 1
  Ul'anovskaya, 198504 St-Petersburg, Russia \\
$^4$ Science Institute, University of Iceland, Dunhagi-3,
  IS-107, Reykjavik, Iceland \\
$^5$ St Petersburg Academic University, 8/3 Khlopina Str, St
  Petersburg, 194021, Russia \\
$^6$ University of Southampton, Highfield, Southampton SO17
  1BJ, United Kingdom
}







\begin{abstract}
We propose a concept of a quantum cascade laser based on transitions of
bosonic quasiparticles (excitons and exciton-polaritons) in a parabolic
potential trap in a semiconductor microcavity. This laser would emit
terahertz radiation due to bosonic stimulation of excitonic transitions.
Dynamics of a bosonic cascade is strongly different from the dynamics of a
conventional fermionic cascade laser. We show that populations of excitonic
ladders are parity-dependent and quantized if the laser operates without an
external terahertz cavity.
\end{abstract}

\maketitle

Quantum cascade lasers (QCLs) are based on subsequent intersubband
transitions of electrons or holes in a Wannier-Stark ladder formed in a
semiconductor superlattice subject to an external electric field~\cite%
{Kazarinov1971,Faist1994,Normand2007}. Emitted terahertz (THz) photons are
polarized perpendicularly to the plane of the structure and propagate
in-plane (which is referred to as wave-guiding or horizontal geometry). QCLs
differ from conventional lasers as they do not require inversion of
electronic population for every particular transition. Still, this is a
fermionic laser, where the quasiparticles emitting light obey Fermi
statistics. Recently, several proposals for bosonic THz lasers based on
exciton-polaritons have been published~\cite{Kavokin2010, Savenko2011}.
These sources are expected to generate THz light beams propagating in the
normal-to-plane direction (vertically) without external THz cavities~\cite%
{Kavokin2012}. The emitted radiation is stimulated by the final state
(exciton-polariton) occupation, which is a purely bosonic effect.

Here we propose a concept of a bosonic cascade laser, which combines
the advantages of QCL (emission of multiple THz photons for each injected
electron) and exciton-polariton lasers (no need for a THz cavity, low
threshold). We consider an exciton cascade
formed by equidistant energy levels of excitons confined in a
parabolic trap in a semiconductor microcavity. Parabolic traps for
exciton-polaritons have been experimentally demonstrated, and an equidistant spectrum of laterally confined exciton-polariton states has been
observed. There are several ways to realize such traps including specially
designed pillar microcavities~\cite{Bajoni2007}, strain induced~\cite{Snoke2007} and optically induced traps~\cite{Amo2010,Wertz2010,Tosi2012}. A particularly promising variation of these designs would be a microcavity with a large parabolic quantum well embedded. We consider the weak coupling regime where the optical mode is resonant with the m$^\mathrm{th}$ exciton level to allow efficient pumping. The other energy levels of the confined excitons would be uncoupled to the cavity mode, forming a dark cascade ideal for a high quantum efficiency device due to the long radiative lifetime. This device would emit radiation polarized in the direction normal to the quantum well plane and propagating in the cavity plane in the wave-guiding regime.
In this Letter we formulate a kinetic theory of bosonic cascade lasers and
calculate the matrix elements of terahertz transitions in realistic
parabolic traps.

The occupation numbers of exciton quantum confined states and the THz
optical mode in our cascade laser can be found from the following set of
kinetic equations ($0$ is the state with the lowest energy, $m$ is the state
with the highest energy, which is resonantly pumped):

\begin{align}
\frac{\mathrm{d}N_{m}}{\mathrm{d}t}& =P-\frac{N_{m}}{\tau }-WN_{m}(N_{m-1}+1)
\notag \\
& \hspace{20mm}+W^{\prime }N_{m-1}(N_{m}+1),  \label{eq:kinetic1} \\
\frac{\mathrm{d}N_{k}}{\mathrm{d}t}& =-\frac{N_{k}}{\tau }+W\left[
N_{k+1}(N_{k}+1)-N_{k}(N_{k-1}+1)\right]   \notag \\
& \hspace{10mm}+W^{\prime }\left[ N_{k-1}(N_{k}+1)-N_{k}(N_{k+1}+1)\right]
\notag \\
& \hspace{25mm}\forall \quad 2\leq k\leq m-1,  \label{eq:kinetic2} \\
\frac{\mathrm{d}N_{0}}{\mathrm{d}t}& =-\frac{N_{0}}{\tau }%
+WN_{1}(N_{0}+1)-W^{\prime }N_{0}(N_{1}+1),  \label{eq:kinetic3} \\
\frac{dn_{\mathrm{THz}}}{dt}& =-\frac{n_{\mathrm{THz}}}{\tau _{\mathrm{THz}}}%
+W\sum_{1}^{m}N_{k}\left( N_{k-1}+1\right)   \notag \\
& \hspace{20mm}-W^{\prime }\sum_{1}^{m}N_{k-1}\left(
N_{k}+1\right).
\end{align}%
Here $W=W_{0}(n_{\mathrm{THz}}+1)$ is the THz emission and $W^{\prime
}=W_{0}n_{\mathrm{THz}}$  is the THz absorption rate, $n_{THz}$ is the THz mode occupation, and $\tau _{\mathrm{THz}}$ is the THz mode lifetime. We assume that the matrix element
of THz transition is non-zero only for neighbouring stairs of the cascade
and that it is the same for all neighbouring pairs. This simplifying assumption
allows for the analytical solution of Eqs. (\ref{eq:kinetic1}) --
(\ref{eq:kinetic3}).  It can be easily relaxed in the numerical calculation,
accounting for the specific conditions of particular experimental systems.

\begin{figure}[h]
\centering
\includegraphics[width=8.116cm]{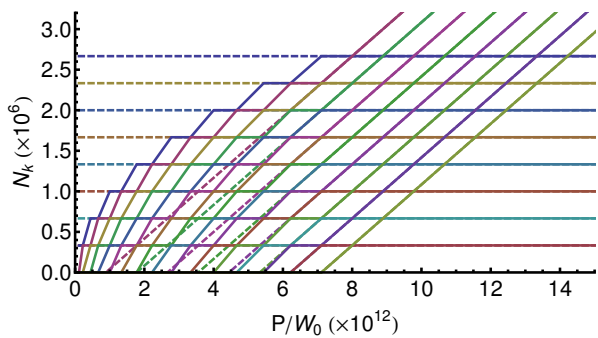}
\caption{(color online) Dependence of the mode occupations in the absence of
a THz cavity on pump intensity, calculated numerically from the kinetic
Eqs.~(\ref{eq:kinetic1}) -- (\ref{eq:kinetic3}) (solid curves)
and analytically from Eqs.~(\ref{eq:An1}), (\ref{eq:An2})
and~(\ref{eq:An3}) (dashed curves). The vertical gray lines correspond
to the step locations given by $P/W_0=n^2/(W_0\protect\tau)^2$ where $n$ is a
half-integer. Parameters: $W_0\protect\tau=3\times10^{-6}$, $m=15$, $n_{\mathrm{THz}}=0$.}
\label{fig:Nk}
\end{figure}

We first consider the case where there is no THz cavity, and assume
that THz photons leave the system immediately such that
$n_{\mathrm{THz}}=0$. The solid curves in Fig.~\ref{fig:Nk} show the
dependence of the mode
occupations on the pump power in this case, which were calculated by
numerical solution of equations~(\ref{eq:kinetic1}) --
(\ref{eq:kinetic3}) for the
steady state. For increasing pump power we see that the modes become
occupied one-by-one and a series of steps appears, each corresponding to the
occupation of an additional mode. In the limit, $W_0\tau\ll1$, the position of
the steps is given by $P/W_0=n^2/(W_0\tau)^2$ where $n$ is a half-integer. For
high pump powers, where all modes are occupied two different behaviours of
the modes can be identified: 0th, 2nd, 4th etc modes continue to increase their
occupation with increasing pump power, while 1st, 3rd, 5th etc modes
have a limited
occupation. This effect persists independently of whether an even or odd
number of modes is considered in the system.

Qualitatively, our results can be understood as follows. Every mode in the
chain experiences both a gain and a loss. The last mode in the chain is
unique since it only experiences loss due to the finite lifetime rather than
THz emission.  Since it experiences loss only due to the lifetime, we can
expect that the last mode is strongly occupied in the limit of high pump
power. This means that the second-to-last mode experiences a strong loss due
to stimulated scattering to a highly occupied state. Thus the second-to-last
mode has a much smaller occupation. The third-to-last mode then experiences
only a small loss due to stimulated scattering and so can again have a large
occupation. The series repeats such that alternate modes have high and low
intensity, with the highly occupied modes introducing a fast loss rate that
limits the occupation of low intensity modes.

Quantitatively, Eqs.~(\ref{eq:kinetic1}) --
  (\ref{eq:kinetic3}) can be solved
analytically in the steady state, where $\sum_k N_k=P\tau$, under the assumption that $W_0\tau\ll1$, $N_{m}\gg1$ and $n_{\mathrm{THz}}=0$. In this regime, $N_0$
depends linearly on the pump intensity:
\begin{equation}
N_0=\frac{P\tau}{\lceil (m+1)/2 \rceil}-\frac{1}{W_0\tau}\lceil\frac{m}{2}\rceil,  \label{eq:An1}
\end{equation}%
where $\lceil n\rceil$ denotes rounding up to the nearest integer. Our approximation makes sense when $N_0$ is positive, i.e.,
when $P\tau>m^2/(2W_0\tau)^2$. The populations of modes with even index also depend linearly on
the pump intensity:
\begin{equation}
N_{2l}=N_{0}+\frac{l}{W_0\tau}\hspace{5mm}\forall\hspace{5mm} 0\leq l\leq \lceil(m-1)/2\rceil
\label{eq:An2}
\end{equation}
The populations of the odd modes are:
\begin{equation}
N_{2l-1}=\frac{l}{W_0\tau}\hspace{5mm}\forall\hspace{5mm} 0\leq l\leq \lceil m/2\rceil\label{eq:An3}
\end{equation}

\begin{figure}[h]
\centering
\includegraphics[width=8.116cm]{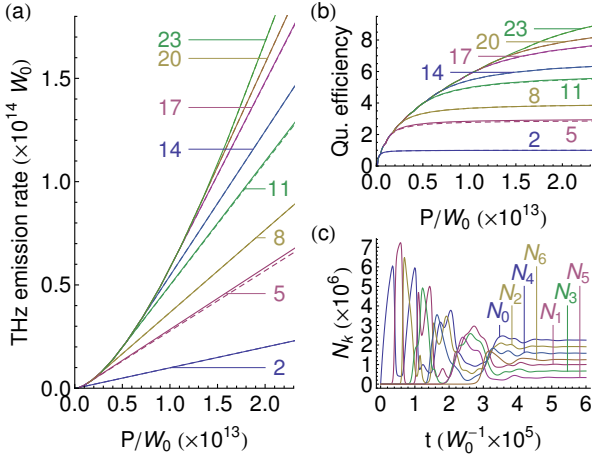}
\caption{(color online) a) Dependence of the THz emission rate on pump
intensity in the absence of a cavity, for different numbers of modes in the
chain (values of $m$ are marked on the plot). Solid curves show results from numerical solution of
Eqs.~(\ref{eq:kinetic1}) -- (\ref{eq:kinetic3}); Dashed
curves show the results of
Eq.~(\ref{eq:AnTHz}), valid for high pump powers. b)
Dependence of the
quantum efficiency on pump intensity (the values of $m$ are the same as in (a)). c) Time dynamics for
$m=6$, $P=3\times10^{12}W$. Parameters: $W_0\tau = 3\times10^{-6}$, $n_{\mathrm{THz}}=0$.}
\label{fig:WTHz}
\end{figure}

Results from Eqs.~(\ref{eq:An1}), (\ref{eq:An2})
  and~(\ref{eq:An3}) are compared
to the numerical results in Fig.~\ref{fig:Nk}. It is also possible to write
an equation for the THz emission rate:
\begin{align}
\frac{dn_{\mathrm{THz}}}{dt}&=W_0\sum_{1}^{m}N_{k}\left(
N_{k-1}+1\right)\\&\approx\frac{m^2+1}{4\tau}\left[N_0-\frac{m-1/2}{3W_0\tau}\right].  \label{eq:AnTHz}
\end{align}%
Results from Eq.~(\ref{eq:AnTHz}) are compared to numerical
calculation of the
THz emission rate in Fig.~(\ref{fig:WTHz})a. For high pump powers, the rate increases linearly with the pump power. This represents a
limit to the quantum efficiency of THz emission, which is given by the THz
emission rate divided by the pump rate, $P$, and plotted in Fig.~\ref%
{fig:WTHz}b. While the presence of the bosonic cascade allows quantum
efficiencies exceeding unity, the quantum efficiency is limited under high
pump powers to $\lceil m/2\rceil $, for $W_0\tau \ll 1$.

Figure~\ref{fig:WTHz}c shows the typical time dependence of the mode
occupations after the pump is switched on (assumed
instantaneously). The enhancement of the scattering via stimulated
processes allows the system to reach equilibrium in a time less than
$1/W_0$. The presence of multiple, dynamically changing effective
scattering rates causes an initially chaotic
dynamics.

\begin{figure}[h]
\centering
\includegraphics[width=8.116cm]{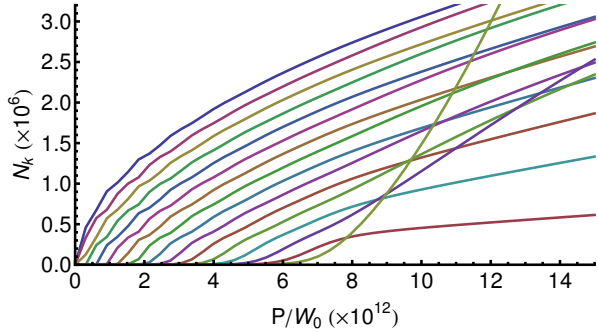}
\caption{(color online) Same as in Fig.~\protect\ref{fig:Nk} with a THz
cavity. $\protect\tau_{THz}=\protect\tau/1000$.}
\label{fig:NkwithTHzCavity}
\end{figure}

\begin{figure}[h]
\centering
\includegraphics[width=8.116cm]{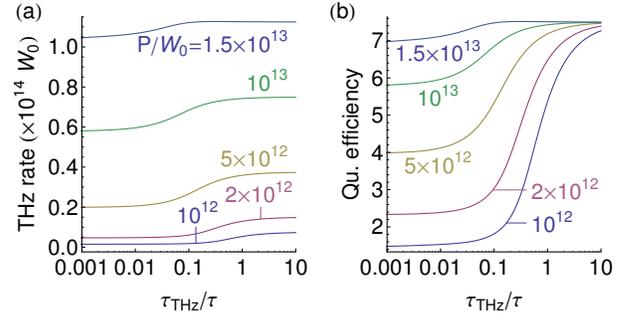}
\caption{(color online) a) Dependence of the THz emission rate in the
presence of a THz cavity on the THz photon lifetime. Different curves correspond to
different pump powers (marked on the plot). b) Dependence of the
quantum efficiency on THz photon lifetime for the same pump powers in (a). Parameters: $m=15$, $W_0\tau = 3\times10^{-6}$.}
\label{fig:WTHzwithTHzCavity}
\end{figure}

In the presence of a THz cavity, macroscopic numbers of THz cavity photons
allow further stimulated enhancement of scattering between the modes. In this case, where upward transitions are allowed in addition to downward ones, the
steps observed in Fig.~\ref{fig:Nk} are washed out as shown in Fig.~\ref{fig:NkwithTHzCavity}. In contrast to Fig.~\ref{fig:Nk}, all modes continue
to increase their population with increasing pump power.
Figure~\ref{fig:WTHzwithTHzCavity}a shows the THz emission rate as a function of the THz photon lifetime. For fixed pump intensity, higher
THz emission rates are observed than in Fig.~\ref{fig:WTHz}a.
Figure~\ref{fig:WTHzwithTHzCavity}b shows the dependence of the THz emission
quantum efficiency on the THz photon lifetime. Increasing of the THz photon
lifetime is able to increase the quantum efficiency, however, not beyond the
limit of $m/2$ already observed in the absence of a THz cavity
for high pump powers.

Let us now demonstrate the feasibility of THz transitions
  between the size-quantized levels of excitons in a parabolic quantum well. The
  two-particle Hamiltonian for an electron and a hole reads
\begin{equation}  \label{Ham}
\mathcal{H }= -\frac{\hbar^2}{2\mu} \frac{\partial^2}{\partial \bm \rho^2} -
\frac{e^2}{\epsilon \rho} -\frac{\hbar^2}{2M} \frac{\partial^2}{\partial \bm %
R^2} + V(z_e,z_h) .
\end{equation}
Here $M=m_e+m_h$, $\mu = m_em_h/M$ is the reduced mass of the
electron-hole pair, $\bm \rho = \bm r_e - \bm r_h = (x,y,z)$ is the relative
coordinate and $\bm R = (m_e \bm r_e + m_h \bm r_h)/M = (X,Y,Z)$ is the center
of mass wave vector, $\epsilon$ is the background dielectric
constant. The quantum well potential $V(z_e,z_h)$ is written in form:
\begin{equation*}
V(z_e,z_h) = A_e z_e^2 + A_h z_h^2,
\end{equation*}
where constants $A_{e,h}$ denote corresponding stiffness. It
  is convenient to rewrite $V(z_e,z_h)$ as a function of the center of
  mass $Z$ and relative motion $z$ coordinates with the result $V =
  V_0(Z,z) + V_1(Z,z)$, where $V_0(Z,z) = (A_e+A_h)Z^2+(m_e^2+m_h^2)z^2/M^2$ does not mix
center of mass and relative motion and
\begin{equation}  \label{V1}
V_1(Z,z) = \frac{2}{M}(m_hA_e - m_eA_h) Zz,
\end{equation}
mixes internal and center of mass degrees of freedom provided that
$m_hA_e \ne m_eA_h$.

In what follows it is assumed that the potential is weak enough,
hence, in the zeroth approximation center of mass can be quantized
independently and $0^{th}$ order wavefunctions have the form:
\begin{equation}  \label{0th}
\Psi_{n,m,h}(\bm r_e, \bm r_h) = R_{nlm}(\rho)
Y_{lm}(\vartheta,\varphi)F_h(Z),
\end{equation}
where $F_h(Z)$ are the eigenfunctions of the potential
$(A_e+A_h)Z^2$, $h$ enumerates levels in this harmonic potential,
$R_{nlm}(\rho)$  are the 3D-hydrogen-like radial functions and
$Y_{lm}(\vartheta,\varphi)$ are (3D) the angular harmonics
of relative motion. $n$, $l$ and $m$ are the quantum numbers of
relative motion of an electron and a hole. Equation~(\ref{0th}) holds
if $\hbar \Omega \ll \mathcal R$, where $\Omega$ is the
eigenfrequency of the center of mass in the parabolic potential,
${M\Omega^2}/{2} = A_e + A_h$, and $\mathcal R = {\mu
  e^4}/({2\epsilon^2\hbar^2})$ is the exciton Rydberg.

It is the mixing of the exciton states caused by
  perturbation Eq.~(\ref{V1}) that gives rise to the THz transitions
  between the neighbouring size-quantized states in a quantum well. Making use of matrix elements of coordinate for harmonic
oscillator and for the hydrogen atom~\cite{Landau1977,ll4_eng} we obtain
for the mixing matrix elements (bra- and ket- denote states as,
e.g. $|n,l,m,h\rangle$) 
\begin{multline}  \label{mixing}
\langle 1,0,0,h-1 |V_1|n, 1,0, h \rangle = \langle 1,0, 0,h |V_1|n, 1,0, h-
1\rangle = \\
\frac{2}{\sqrt{3}M}(m_eA_e-m_hA_h) a_0 f_n \sqrt{\frac{h\hbar}{2M\Omega}}
\equiv V_h(n),
\end{multline}
where $a_0 = \frac{\hbar^2\epsilon}{\mu e^2}$ is the exciton
  Bohr radius and
\begin{equation*}
f_n = \frac{2^4(n-1)^{n-5/2}n^{7/2}}{(n+1)^{n+5/2}}.
\end{equation*}
Using first order perturbation theory, the THz
  transition dipole moment between the neighbouring states is:
\begin{equation}  \label{Dh}
D_h = e\langle \tilde \Psi_{h-1} | x| \tilde \Psi_{h} \rangle = e
\sum_{n,\pm} \frac{V_h(n)}{\Delta_{h,\pm}(n)} \frac{f_n a_0}{\sqrt{3}},
\end{equation}
where the wavefunctions of $1s$ exciton states take the form in the leading
order in $V_1$:
\begin{multline}  \label{Psi}
\tilde{ \Psi}_{1,0,0,h}(\bm \rho, \bm R) = R_{100}(\rho)
Y_{00}(\vartheta,\varphi) F_h(Z) + \\
\sum_{n,\pm} \frac{V_h(n)}{\Delta_{h,\pm}(n)} R_{n10}(\rho)
Y_{10}(\vartheta,\varphi) F_{h\pm 1}(Z).
\end{multline}
Here $\Delta_{h,\pm}(n) = E_{1s} - E_{np} \mp \hbar \Omega
  \approx E_{1s}
  - E_{np}$, and $E_{ns} = E_{np} = - {\mathcal R}/{n^2}$.
Estimation according to Eq.~(\ref{Dh}) shows for $a_0=100$~\AA,
$\mathcal R = 5$~meV,
$\hbar\Omega=1$~meV, $m_e = 0.1m_0$, $m_h=0.4m_0$ and $A_e = A_h$
that
\begin{equation}  \label{est}
|D_h| = e \sqrt{h} \times 8.3\times 10^{1}~\mbox{\AA},
\end{equation}
which is comparable with $2p\to 1s$ transition strength. Following Ref.~\cite{Kavokin2010}, this gives the transition rate for $1$THz emission in the kinetic equations $W_0\approx5500$s$^{-1}$ (where we took the refractive index as $3$ corresponding to a GaAs based system). For excitons with lifetime $\sim500$ps, this gives $W_0\tau\approx3\times10^{-6}$.

\begin{figure}[h]
\centering
\includegraphics[width=8.116cm]{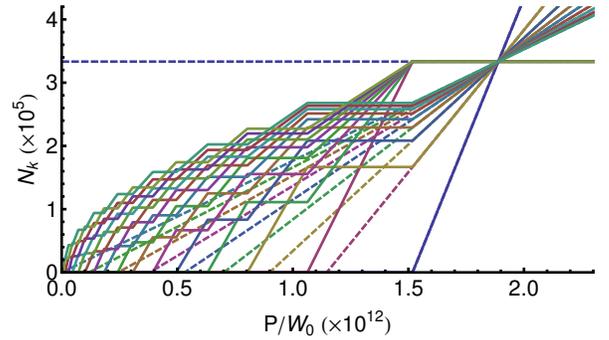}
\caption{(color online) Same as in Fig.~\protect\ref{fig:Nk} with transition rates proportional to the mode index.}
\label{fig:Wpropton}
\end{figure}

Since the transition rate is proportional to $|D_h|^2$,
Eq.~(\ref{est}) introduces a proportionality between the
transition rate and the mode indices, $n$, for the specific case of a
parabolic trap with electron-hole mixing. This can be modelled by
straightforward modification of the rate
Eqs.~(\ref{eq:kinetic1}) -- (\ref{eq:kinetic3}). While this
can distort
the steps observed in Fig.~\ref{fig:Nk} as shown in
Fig.~\ref{fig:Wpropton}, the behaviour is similar. For $W_0\tau\ll1$
and $m$ even, one can obtain analytically:
\begin{equation}
N_{2k}\approx\frac{1}{W_0\tau}-\frac{m\left(1+m-PW_0\tau^2\right)\Gamma(1/2+k)
  \Gamma(m/2)}{4W_0\tau\Gamma(1+k)\Gamma[(3+m)/2]},
\end{equation}
where $k$ is an integer and $\Gamma(n)$ is the Euler gamma function. The
other modes of the system all have the same intensity,
$N_{2k+1}\approx1/(W_0\tau)$ and the THz emission rate is given by
$\left(2P\tau-m/(W_0\tau)-N_0+N_m\right)/(W_0\tau)$.

Finally, we note that the waveguiding geometry allows for the increase of the active
volume of the laser and, consequently, the output power. Its drawback is a
larger size compared to the vertical cavity laser geometry. Lateral polariton
traps would allow a more compact size of the device, while the matrix elements
of terahertz transitions are believed to be significantly larger in
parabolic quantum wells.

In summary, we have proposed the use of a bosonic cascade to implement THz
lasing with high quantum efficiency, above unity. Such a device requires a
parabolic trapping, which can be arranged in the plane of a semiconductor
microcavity via a variety of techniques or using parabolic wide quantum
wells. Electron-hole mixing was shown to give rise to non-zero transition
matrix elements. A series of steps in the mode occupations of the bosonic
cascade was predicted for increasing pump power in the absence of a THz
cavity. The quantum efficiency of a device can be improved by increases in
the pump power or the use of a THz cavity, however, the quantum efficiency
is limited to one-half of the number of modes present.

This work has been supported by the EU IRSES projects \textquotedblleft POLAPHEN \textquotedblright and \textquotedblleft POLATER \textquotedblright. IAS
acknowledges the support from Rannis \textquotedblleft Center of Excellence
in Polaritonics\textquotedblright. MMG was partially
  supported by RFBR and RF President Grant NSh-5442.2012.2.


\begin{thebibliography}{99}
\bibitem{Kazarinov1971} R F Kazarinov \& R A Suris, Sov. Phys.
Semiconductors, \textbf{5}, 707 (1971).

\bibitem{Faist1994}
J Faist, F Capasso, D L Sivco, C Sirtori, A L Hutchinson, \& A Y Cho, Science, \textbf{264}, 553 (1994).

\bibitem{Normand2007} E Normand \& I Howieson, Laser Focus World, \textbf{43}%
, 90 (2007).

\bibitem{Kavokin2010} K V Kavokin, M A Kaliteevski, R A Abram, A V Kavokin, S Sharkova, \& I A Shelykh, Appl. Phys. Lett. \textbf{97}, 201111 (2010).

\bibitem{Savenko2011} I G Savenko, I A Shelykh, \& M A Kaliteevski, Phys.
Rev. Lett., \textbf{107}, 027401 (2011).

\bibitem{Kavokin2012} A V Kavokin, I A Shelykh, T. Taylor \& M M Glazov, Phys. Rev.
Lett., \textbf{108}, 197401 (2012).

\bibitem{Bajoni2007} 
D Bajoni, et al., Appl. Phys. Lett., \textbf{90}, 051107 (2007).

\bibitem{Snoke2007} R Balili, V Hartwell, D Snoke, L Pfeiffer, \& K West, Science, \textbf{316}, 1007 (2007).

\bibitem{Amo2010} 
A Amo, et al., Phys. Rev. B, \textbf{82}, 081301(R) (2010).

\bibitem{Wertz2010} 
E Wertz, et al., Nature Phys., \textbf{6}, 860 (2010).

\bibitem{Tosi2012} 
G Tosi, et al., Nature Phys., \textbf{8}, 190
(2012).

\bibitem{Landau1977} L Landau \& E Lifshitz, \textit{Quantum Mechanics:
Non-Relativistic Theory (vol. 3)}, Butterworth-Heinemann, Oxford (1977).

\bibitem{ll4_eng}
V B Berestetskii, L P Pitaevskii, \& E M Lifshitz, {\it Quantum Electrodynamics, Second Edition (vol. 4)}, Butterworth-Heinemann, Oxford, (1999).
\end{thebibliography}
\end{document}